\documentclass[preprint,showpacs,preprintnumbers,amsmath,amssymb,superscriptaddress,12pt,graphicx]{revtex4}

\usepackage{graphicx}
\usepackage{dcolumn}
\usepackage{bm}
\usepackage{amssymb}

\renewcommand{\L}{{\cal L}}

\newcommand{\be}{\begin{equation}}
\newcommand{\ee}{\end{equation}}
\newcommand{\bee}{\begin{eqnarray}}
\newcommand{\een}{\end{eqnarray}}
\newcommand{\ba}{\begin{eqnarray}}
\newcommand{\ea}{\end{eqnarray}}

\begin{document}

\title{Evolution of magnetic component in Yang-Mills condensate dark energy models }

\author{Wen Zhao}
\email{wzhao7@mail.ustc.edu.cn}\affiliation{Department of Applied Physics, Zhejiang University of Technology,  Hangzhou, Zhejiang}\affiliation {Department of Physics and Astronomy, Cardiff University, Cardiff, United Kingdom} \affiliation{Wales Institute of Mathematical and Computational Science, Swansea, United Kingdom} 
\author{Donghui Xu}
\affiliation{Department of Applied Physics, Zhejiang University of Technology,  Hangzhou, Zhejiang}



\begin{abstract}
The evolution of the electric and magnetic components in
an effective Yang-Mills condensate dark energy model is
investigated. If the electric field is dominant, the magnetic
component disappears with the expansion of the Universe. The total
YM condensate tracks the radiation in the earlier Universe, and
later it becomes $w_y\sim-1$ thus is similar to the cosmological
constant. So the cosmic coincidence problem can be avoided in this
model. However, if the magnetic field is dominant, $w_y>1/3$ holds
for all time, suggesting that it cannot be a candidate for the
dark energy in this case.
\end{abstract}



\maketitle


\section{Introduction}

Recent observations on the Type Ia Supernova\cite{sn}, Cosmic
Microwave Background Radiation\cite{map} and Large Scale
Structure\cite{sdss} all suggest a flat Universe consisting of
dark energy (73\%), dark matter (23\%) and baryon matter (4\%). It
is important to understand the physics of the mysterious dark
energy, which has the equation of state (EOS) $w<-1/3$ and causes
the recent accelerating expansion of the Universe. The simplest
model is the cosmological constant $\Lambda$ with
$\omega_{\Lambda}\equiv-1$, which fits the observation fairly
well. However, a number of evidences suggest that the EOS of the
dark energy may evolve. This has stimulated a number of approaches
to build the dark energy models with a dynamic field. One class of
approaches is to introduce a scalar field, such as the
quintessence\cite{quint}, the phantom\cite{phantom}, the
k-essence\cite{k} and the quintom\cite{quintom}. Another class of
models is based on the conjecture that a vector field can be the
origin of the dark energy\cite{vector}, which has different
features to those of scalar field. In the
Ref.\cite{z,Zhang,zhao,coincidence,couple}, it is suggested that
the Yang-Mills (YM) field can be a kind of candidate for such a
vector field.

Compared with the scalar field, the YM field is the indispensable
cornerstone to particle physics and the gauge bosons have been
observed. There is no room for adjusting the form of effective YM
Lagrangian as it is predicted by quantum corrections according to
field theory. In the previous works, we have investigated the
simplest case with only electric component and found attractive
features: 1) this dark energy can naturally get the EOS of
$w_y>-1$ and $w_y<-1$\cite{zhao}, which is different from the
scalar quintessence models; 2) with the expansion of the Universe,
the EOS of the YM condensate naturally turns to the critical state
of $w_y=-1$\cite{zhao}, consistent to the
observations\cite{seljak}; 3) the cosmic coincidence problem is
naturally avoided in the YM condensate dark energy
models\cite{coincidence, couple}; 4) the EOS of the dark energy
can cross $-1$ in the double-field models or coupled
models\cite{zhao,couple}; 5) the big rip is naturally avoided in
the models\cite{couple}.

In this letter, we discuss the evolution of the YM condensate dark
energy with both electric and magnetic components. We find that,
if the magnetic component is subdominant in the initial condition,
it rapidly decreases to zero with the expansion of the Universe.
The states of $w_y>-1$ and $w_y<-1$ all can be realized in the
models. In the former case, the state of YM condensate is
$w_y\sim1/3$ in the earlier stage, and later it turns into
$w_y\sim-1$, which is similar to the case with only electric
component. So the cosmic coincidence problem is naturally avoided
in the models. However, if the magnetic component is dominant in
the initial condition, the state of YM condensate keeps $w_y>1/3$,
which cannot be a candidate for dark energy.


\section{The effective Yang-Mills field model}

The effective Lagrangian density of the YM field up to 1-loop
order is\cite{pagels, adler}
 \be
 \L_{eff}=\frac{b}{2}F\ln\left|\frac{F}{e\kappa^2}\right|.
 \ee
Here $b=11N/24\pi^2$ for the generic gauge group $SU(N)$ is the
Callan-Symanzik coefficient\cite{Pol},
$F=-(1/2)F^a_{\mu\nu}F^{a\mu\nu}$ plays the role of the order
parameter of the YM condensate, and $\kappa$ is the
renormalization scale with the dimension of squared mass which is
the only model parameter. This effective YM Lagrangian exhibits
the features of the gauge invariance, the Lorentz invariance, the
correct trace anomaly, and the asymptotic freedom\cite{pagels}.
With the logarithmic dependence on the field strength, $\L_{eff}$
has a form similar to the Coleman-Weinberg scalar effective
potential\cite{coleman} and the Parker-Raval effective gravity
Lagrangian\cite{parker}. The effective YM condensate was firstly
put into the expanding Friedmann-Robertson-Walker (FRW) spacetime
to study inflationary expansion in the Ref.\cite{z} and the dark
energy in the Ref.\cite{Zhang}.  Following the
Refs.\cite{zhao,coincidence}, we work in a spatially flat FRW
spacetime with a metric
 \be
 ds^2=a^2(\tau)(d\tau^2-\delta_{ij}dx^idx^j),\label{me}
 \ee
where $\tau=\int(a_0/a)dt$ is the conformal time. Assume that the
Universe is filled with the YM condensate. For simplicity we study
the $SU(2)$ group. The energy density and pressure are given by

 \be
 \rho_y=\frac{1}{2}b\varepsilon(E^2+B^2)+\frac{1}{2}b(E^2-B^2),
 \ee
 \be
 p_y=\frac{1}{6}b\varepsilon(E^2+B^2)-\frac{1}{2}b(E^2-B^2),
 \ee
where the dielectric constant is given by $b\varepsilon\equiv
b\ln|(E^2-B^2)/\kappa^2|$. We define two dimensionless quantities
$f\equiv (E^2-B^2)/\kappa^2$ and $q\equiv (E^2+B^2)/\kappa^2$. It
is easy to find that $q\ge f$, and $q=f$ only if $B^2=0$. The
energy density and pressure can be rewritten as
 \be
 \rho_y=\frac{1}{2}b\kappa^2(\varepsilon q+f),~~~~p_y=\frac{1}{2}b\kappa^2\left(\frac{1}{3}\varepsilon
 q-f\right).\label{rhoy}
 \ee
The energy density of YM condensate should has the positive value,
which follows a constraint of the YM condensate
 \be\label{ym1_constraint}
 \varepsilon q+f~>~0.
 \ee
The EOS of the YM condensate is
 \be
 w_y=\frac{\varepsilon q-3f}{3\varepsilon q+3f}.\label{om}
 \ee
At the critical point with $|f|=1$, one has $\varepsilon=0$ and
$w_y=-1$. Around this critical point, $|f|<1$ gives
$\varepsilon<0$ and $\varepsilon<-1$, and $|f|>1$ gives
$\varepsilon>0$  and $w_y>-1$. So in the YM field model, EOS of
$w_y >-1$ and $w_y<-1$ all can be naturally realized.

The effective YM equations are\cite{zhao,coincidence}
 \be
 \partial_{\mu}(a^4\varepsilon F^{a\mu\nu})+f^{abc}A_{\mu}^{b}(a^4\varepsilon
 F^{c\mu\nu})=0,
 \ee
which can be reduced to
 \be
 \partial_{\tau}(a^2 \varepsilon E)=0.
 \ee
At the critical point with $\varepsilon=0$, this equation is an
identity. And when $\varepsilon\neq0$, this equation has an exact
solution
 \be\label{ym1_2}
 q+f=c~a^{-4}\varepsilon^{-2},
 \ee
where $c$ is the integral constant, and $q$ and $f$ are the
variables. The energy conservation equation of the YM condensate
is
 \be
 \left(a^3\rho_y\right)'=-p_y,
 \ee
where the prime denotes $d/d(a^3)$. This equation can be reduced
to
 \be\label{ym1_3}
 \left(1+\frac{q}{f}\right)f'+\varepsilon
 q'=-\frac{4}{3}\varepsilon qa^{-3}~.
 \ee
By the equations of (\ref{ym1_2}) and (\ref{ym1_3}), one can
numerically solve the evolution of the EOS of the YM dark energy.
It is easily to find that, in the YM condensate models, the
conformal time $\tau$ can be entirely replaced by the scale factor
$a$ and the evolution of the YM condensate with the scale factor
is independent of the other components in the Universe. In the
following discusses, we choose the initial condition at $a=a_i$,
where $a_i$ can be at any time, and the initial condition is
chosen as
 \be
 q~=~q_i,~~~~~f~=~f_i.
 \ee
So the integral constant $c$ is fixed
 \be
 c=(q_i+f_i)(\ln f_i)^2.
 \ee

First, we consider the case of the electric field dominant in the
initial condition, which requires an inequality
 \be
 f_i>0.
 \ee
The value of $w_i$ is exactly determined by the values of $f_i$
and $q_i$. Here we consider two kinds of choices of the initial
condition, $w_i>-1$ and $w_i<-1$.

The initial condition of $w_i>-1$ requires
 \be
 q_i~>~f_i~>~1.
 \ee
The value of $q_i$ closer to $f_i$ suggests that the density of
electric field is much larger than which of magnetic field, and
$q_i=f_i$ suggests that the YM condensate includes only electric
component. When the values of $q_i$ and $f_i$ are all close to
$1$, it means that $E^2\rightarrow\kappa^2$ and $B^2\rightarrow0$.
On the contrary, the value of $q_i$ much larger than $f_i$
suggests that the density of electric field is much closer to that
of magnetic field. Here we consider three different models:

{\bf Mod.a1}: $f_i=50$,~$q_i=100$; {\bf Mod.a2}:
$f_i=5$,~$q_i=100$; {\bf Mod.a3}: $f_i=5$,~$q_i=10$.

Solving the Eqs.(\ref{ym1_2}) and (\ref{ym1_3}), we get the
evolution of the EOS of the YM condensate in these models, which
are plotted in Fig.1. We find that the evolution of the EOS is
similar in all these models: in the earlier stage, $w_y\sim 1/3$,
tracking the evolution of the radiation, and the energy density
$\rho_y\propto a^{-4}$. However at a transition time, $w_y$
rapidly transits from $w_y\sim1/3$ to an attractor solution of
$w_y\sim-1$, similar to the cosmological constant, and energy
density of YM condensate keeps constant. This feature is same with
the simple YM dark energy model with only electric
field\cite{coincidence}. As is known, an effective theory is a
simple representation for an interacting quantum system of many
degrees of freedom at and around its respective low energies.
Commonly, it applies only in low energies. However, it is
interesting to note that the YM condensate model as an effective
theory intrinsically incorporates the appropriate states for both
high and low temperature. As has been shown above, the same
expression in Eq.(\ref{om}) simultaneously gives $p_y\rightarrow
-\rho_y$  at low energies, and $p_y\rightarrow \rho_y/3$  at high
energies. Therefore, our model of effective YM condensate can be
used even at higher energies than the renormalization scale
$\kappa$.

Fig.2 plots the evolution of electric and magnetic components in
these three models. We find that their evolution processes are
different in these models. For the magnetic component, in the
earlier stage, $B^2\propto a^{-4}$, and after the transition time
(which is also the transition time of EOS from $w_y\sim1/3$ to
$w_y\sim-1$), the values of $B^2$ rapidly decrease to zero. For
the electric component, in the earlier stage, the value of $E^2$
is also $\propto a^{-4}$, but after the transition time, the value
of $E^2$ stops decreasing and approaches to the critical state of
$E^2=\kappa^2$, the renormalization scale. The electric component
dominates in the YM condensate in all time. If in the end of the
reheating, a very early stage of the Universe, the energy density
of the YM condensate is smaller, corresponding to a smaller $E^2$,
it decreases as $E^2\propto a^{-4}$ and arrives at the state of
$E^2\sim\kappa^2$ earlier, and the transition time is also
earlier. On the contrary, a larger $E^2$ in the very early
Universe leads to a latter transition time. So the transition time
of the EOS of the YM condensate is directly determined by the
choice of initial condition in the very early Universe. However,
no matter what initial condition one chooses, the YM condensate
must arrive at the attractor solution of $E^2\rightarrow\kappa^2$,
$B^2\rightarrow0$ and $w_y\rightarrow-1$. In this solution, the
energy density of YM condensate is
 \be
 \rho_y\rightarrow\frac{b\kappa^2}{2},
 \ee
which is independent of the choice of the initial condition. So
the cosmic coincidence problem is naturally avoided. In order to
account for the present observational value of the dark energy,
one needs to finely tune the value of the renormalization scale
$\kappa\simeq3.57h\times10^{-5}eV^2$\cite{coincidence}, where $h$
is the Hubble constant. This energy scale is low compared to
typical energy scales in particle physics. So the ``fine-tuning"
problem is present in these models. From these models, we also
find that the EOS of the YM condensate cannot cross $-1$, which is
same with quintessence models\cite{quint}, unless the coupling of
the YM condensate with the matter is considered\cite{couple}.

Now we turn to another case of $w_i<-1$. Eq.(\ref{ym1_constraint})
requires that
 \be
 f_i<q_i<-f_i/\ln f_i~,~~~~0<f_i<1,
 \ee
which leads to the constraint of $e^{-1}<f_i<1$. Here we also
consider three different models:

{\bf Mod.b1}: $f_i=0.9$,~$q_i=2.0$; {\bf Mod.b2}:
$f_i=0.9999$,~$q_i=10.0$; {\bf Mod.b3}: $f_i=0.5$,~$q_i=0.6$.

In Fig.3, we plot the evolution of the EOS of the YM condensate in
these three models, and Fig.4 plots the evolution of the electric
and magnetic components. Similar to the previous three models,
with the expansion of the Universe, the EOS of the YM condensate
runs to the critical state of $w_y=-1$, the density of the
electric field approaches to the value of $\kappa^2$, which is
dominant in YM condensate in all time, and the energy density of
the magnetic field approaches to zero. The total energy density of
the YM condensate $\rho_y\rightarrow b\kappa^2/2$, and the
``fine-tuning" problem also exists. From these models, we also
find that the EOS of the YM condensate cannot cross $-1$, which is
same with phantom models\cite{phantom}.

From these discussion, we get the conclusions: if the electric
component dominates in the initial condition, the YM condensate
can have the state of $1/3>w_y>-1$ or $w_y<-1$, depended on the
choice of the initial condition. The former is similar to the
quintessence models, and the cosmic coincidence problem is
naturally avoided. The latter is similar to the phantom models. In
each case, the EOS of the YM condensate approaches to $-1$ in the
latter stage, similar to the cosmological constant, which is
independent of the choice of the initial condition. The value of
$E^2$ approaches to $\kappa^2$, the renormalization scale, and the
value of $B^2$ approaches to zero.

Second, we consider the case of magnetic field dominant. In the
extreme condition of $E^2=0$, the energy density, pressure and the
EOS of the YM condensate are
 \be
 \rho_y=\frac{1}{2} b
 B^2(\varepsilon-1),~~p_y=\frac{1}{2} b
 B^2\left(\frac{1}{3}\varepsilon+1\right),~~w_y=\frac{\varepsilon+3}{3\varepsilon-3},
 \ee
respectively, where $\varepsilon=\ln(B^2/\kappa^2)$. The
constraint of $\rho_y>0$ is reduced to
 \be\label{theta}
 \varepsilon>1,
 \ee
which leads to a constraint of the EOS of the YM condensate
 \be
 w_y>\frac{1}{3}~.
 \ee
This means that this YM condensate cannot get a negative pressure,
and be a candidate for dark energy. We turn to the general case
with the magnetic component dominant. The energy density, pressure
and EOS of the YM condensate are in the Eqs.(\ref{rhoy}) and
(\ref{om}). The constraint of $\rho_y>0$ yields
 \be
 0>f>-\varepsilon q,
 \ee
which follows that $w_y>1/3$, The YM condensate cannot get a
negative EOS. However, if it is possible for the YM condensate to
evolute from the state of $B^2>E^2$ to the state of $B^2<E^2$, and
get a negative pressure? If this is possible, in the transition
point, the YM condensate must have a state of $B^2=E^2$, where
$\varepsilon=\infty$. From the effective YM equation
(\ref{ym1_2}), one knows this only occurs at $a=0$. So the
transform from the state of $E^2>B^2$ to $E^2<B^2$ cannot realize.
In conclusion, the YM condensate with magnetic component dominant
cannot get a negative pressure, so it cannot be a candidate for
dark energy.

\section{Summary}

The evolution of the YM condensate as a candidate for dark energy
is investigated, which has no free parameters except the value of
the present cosmic energy scale, and the cosmic evolution entirely
depends on the initials condition. This study shows that the
evolution of the electric and magnetic components in the YM
condensate is different for the models with different initial
conditions. If the electric component is dominant in the initial
condition, and $w_i>-1$ is satisfied, $E^2\propto a^{-4}$ in the
earlier stage, and later it turns to the state of
$E^2\rightarrow\kappa^2$. For the magnetic component, $B^2\propto
a^{-4}$ in the earlier stage, and later it decreases rapidly to
zero. The electric component is dominant in the YM condensate in
all time, and the total EOS of the YM condensate transits from the
state of $w_y\sim1/3$ to the state of $w_y\sim-1$. So the cosmic
coincidence problem is naturally avoided in the models. If in the
initial condition, the electric component is dominant and $w_i<-1$
is satisfied, the electric component runs to the state of
$E^2\rightarrow\kappa^2$ and the magnetic component runs to
$B^2\rightarrow0$ in the later stage of the Universe. The total
EOS of the YM condensate keeps $w_y<-1$, and later it turns to a
state of $w_y\rightarrow-1$. So the big rip problem is avoided.
However, if the magnetic component is dominant in the initial
condition, $w_y>1/3$ is satisfied for all time. So it cannot be a
candidate for dark energy.


\section*{Acknowledgements}
W.Zhao thanks Y.Zhang for help discussions.
This work is supported by CNSF No. 10703005 and 10775119, the
Research Funds Launched in ZJUT No.109001729.


\newpage

\begin{figure}[t]
\centerline{\includegraphics[width=10cm]{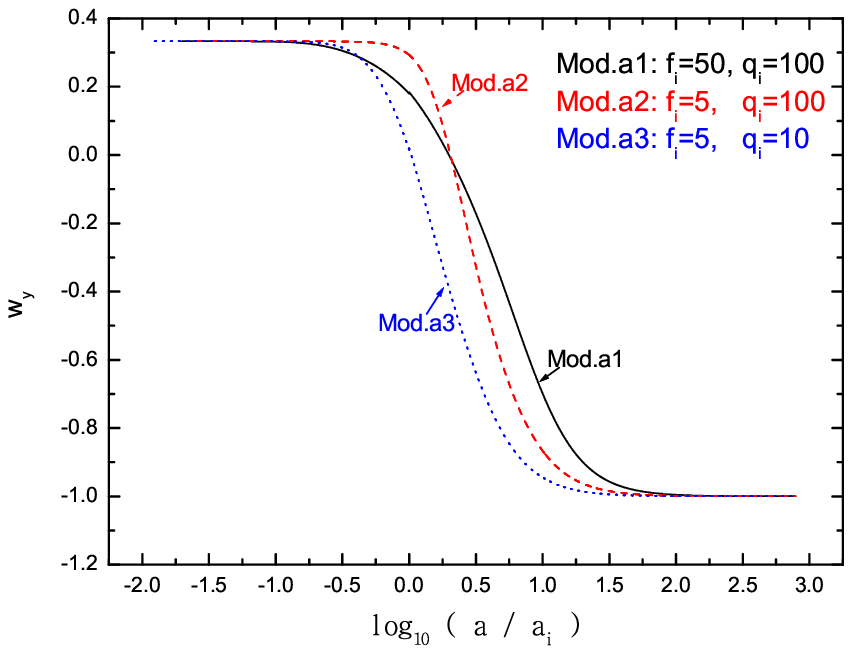}}
\caption{In
models a1,~a2,~a3, the evolution of the EOS of the YM condensate
with the scale factor $a$.}\label{2411}
\end{figure}

\begin{figure}[t]
\centerline{\includegraphics[width=10cm]{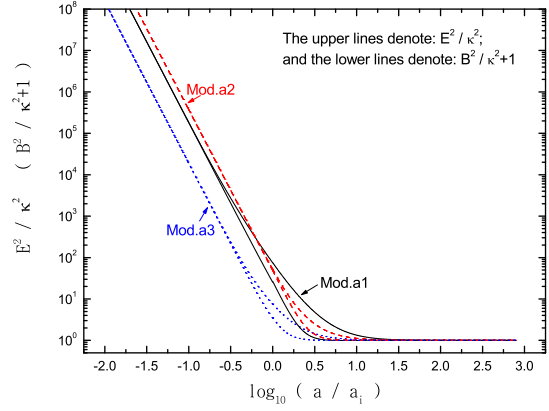}}
\caption{In
models a1,~a2,~a3, the evolution of the ``electric" and
``magnetic" components with the scale factor $a$.}\label{2421}
\end{figure}

\begin{figure}[t]
\centering \centerline{\includegraphics[width=10cm]{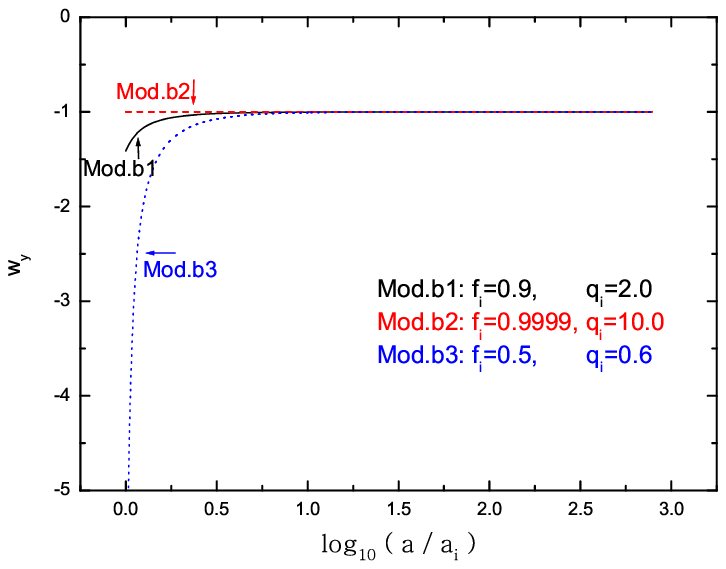}}
\caption{In models b1,~b2,~b3, the evolution of the EOS of the YM
condensate with the scale factor $a$.}\label{2431}
\end{figure}

\begin{figure}[t]
\centerline{\includegraphics[width=10cm]{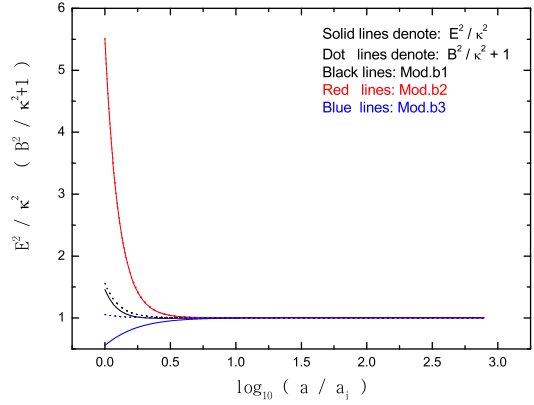}}
\caption{In
models b1,~b2,~b3, the evolution of ``electric" and ``magnetic"
component with the scale factor $a$.}\label{2441}
\end{figure}

\end{document}